# Amorphous carbon films in direct current magnetron sputtering from regenerative sooting discharge

**\*Sumera Javeed[1,2], Sumera Yamin[1], Sohail Ahmad Janjua[1], Kashif Yaqub[1,2], Afshan Ashraf[1,2], Sumaira Zeeshan[1,2], Mazhar Mehmood[2], Muhammad Anwar-ul-Haq[1], and Shoaib Ahmad[1]**

[1] Accelerator and Carbon Based Nanotechnology Laboratory, PINSTECH

[2] Pakistan Institute of Engineering and Applied Sciences (PIEAS)

P. O. Box Nilore, Islamabad, Pakistan.

## Abstract

We present results of carbon coatings on metal substrates in cylindrical hollow cathode (CHC) direct current magnetron sputtering. This is a new technique of making amorphous carbon film in CHC magnetron sputtering from regenerative sooting discharge. The carbon films are deposited on Cu and Al substrates in Ne atmosphere and compared with the films of carbon soot on the same materials produced from conventional arc discharge between graphite electrodes at 80 A in He background. The films are characterized using online emission, Raman, and Fourier transform infrared spectroscopy; X-ray diffraction (XRD) and Scanning electron microscopy (SEM). Raman spectroscopy reveals the existence of graphite and diamond like structures from arc discharge while in CHC magnetron sputtering, graphite like structures are dominant. XRD pattern from arc discharge show precipitates of $Al_4C_3$ of rhombohedral and hexagonal types in nanometer ranges for aluminum sample and probable formation of diamond and hexagonal carbon in copper whilst in magnetron sputtering we get amorphous carbon films. SEM images of surface show collection of loose agglomerates of carbon particles in arc discharge whereas for magnetron sputtering structures are regular with smooth edges and fine grains.

**\*Corresponding author: Tel. +92-51-2207276, Fax. +92-51-9248808

E-mail address: subahyaqeen@gmail.com

**Introduction:**

A new technique of carbon coatings on metal substrates in cylindrical hollow cathode (CHC) direct current (DC) magnetron sputtering in regenerative sooting discharge is presented. The results are compared with continuous arc, which is non-regenerative sooting discharge i.e. soot is produced and deposited on substrates without recycling. On the other hand, the CHC magnetron sputtering is operated in regenerative sooting discharge mode in which kinetic and potential sputtering introduces carbonaceous environment containing all sorts of carbon nanoparticles that are settled on the cathode surface and a layer of soot is formed. The collisions of carbon nanoparticles with energetic electrons and with other constituents of discharge cause this soot to regenerate [1,2]. Therefore, there is a subtle difference in the production of soot in such a regenerative environment and the one where sublimation dominates (non-regenerative mode). In continuous arc discharge removal of surface constituents of graphite anode is characterized by sp2 bond breaking due to thermal energy provided by electrons. Emission of monatomic, diatomic and larger carbon clusters $C_x$ ($x \geq 3$) occurs due to the sublimation of graphite anode. $C_2$ has unique properties and plays an important role in the formation and fragmentation processes leading to closed cage carbon clusters. Our theoretical studies have highlighted the role of $C_2$ in self-assembly of fullerenes and single walled nanotubes [3,4].

The continuous arc discharge method described here is similar to that of Krätschmer et al [5] and was also recently employed to study the sublimation of graphite [6]. Arc discharges are operated from atmospheric pressures down to about 250 Torr with voltage of about 20 V and current up to 80 A [7]. Helium is selected due to its better thermal conductivity.

The magnetron sputtering developed and reported here has got many special features. The most prominent is the type of discharge governed by regenerative soot. Other feature is placement of metallic substrate inside the graphite CHC having special carbonaceous environment along with application of axial magnetic field perpendicular to metallic sample. The films are formed from deposition of neutral carbon nanoparticles, negatively charged carbon ions as well as metastable carbon particles. Magnetic field not only stabilizes the plasma at lower pressure but also enhances sputtering yield that helps to acquire the required thickness in relatively shorter time. Substrate is additionally bombarded by electrons affected by presence of electric (E) and magnetic fields (B); electrons moving perpendicular to surface are gyrated due to B and electrons that are parallel to surface but perpendicular to magnetic field are deflected towards the surface due to E. Ne is selected because it has lower ionization potential and we need to ionize the gas to get sufficient number of electrons available to sputter cathode surface.

Cu and Al are selected as substrates to deposit carbon film due to their entirely different properties with reference to their interaction with carbon. Cu is almost non reactive due to miscibility gap whereas Al strongly reacts with carbon. Many researchers have reported use of Cu and Al substrates for formation of nanostructures. Carbon ions were implanted into Cu and Al for studying growth of carbon nanostructures [8,9].

DC magnetron sputtering is extensively used for film formation due to its ease of operation and flexibility for its selection of parameters. Many researchers used DC magnetron sputtering system to study different physical and chemical properties of amorphous carbon films under various experimental conditions. Cho et al studied chemical structure and physical properties that is electrical resistivity, mass density, hardness and elastic modulus of amorphous carbon films deposited on NaCl, Si wafer and Silica glass. Results were interpreted as a function of sputtering power density [10]. Mounier et al investigated amorphous carbon films and studied the growth rate, composition, electrical resistivity, mass density, refractive index and microstructure of the films as a function of substrate temperature in the range of 50-350 °C [11]. Void distribution in amorphous carbon films deposited on pdoped silicon crystals and onto ultra pure Al foils were investigated by Freire et al. They concluded that films deposited at lower plasma pressure are hard and dense; however, soft films grown at higher pressure have open microstructures [12].

Raman spectra of a wide range of disordered and amorphous carbons were measured under excitation from 785 to 229 nm by Ferrari and Robertson. The model and theoretical understanding was also studied [13,14]. An important property of interest in carbon coatings is their relative $sp^2/sp^3$ bonding ratio, which can be related to the type of structure present. Zhang et al used electron energy loss spectroscopy to quantify $sp^2/sp^3$ bonding in the amorphous carbon coatings of less than 100 nm deposited on KBr and Si wafer substrates [15]. Hard amorphous carbon coatings have been studied using magnetron sputtering and very high toughness films were produced. A relationship between surface roughness, hardness, adhesion to substrate and coefficient of friction relative to $sp^2/sp^3$ ratio is discussed [16]. Paul et al used this ratio to study hydrophobicity of diamond like carbon films and interpreted results in terms of hybridization [17]. Recently CHC magnetron sputtering system was used by Duarte et al to deposit crystalline titanium dioxide thin film on p-silicon substrate [18].

## 2. Experimental

Aluminum and copper substrates are placed inside the discharge chamber to get carbon films as a result of soot deposition in both techniques. Metallic substrates are fine polished and cleaned prior to coating. Substrates were not externally heated.

### 2.1 Continuous Arc Discharge

Fig 1 shows the schematic of the experimental setup used for film formation as a result of soot deposited on metal substrates in continuous arc discharge. Between graphite electrodes 20 V and 80 A is applied in the presence of He gas at 250 Torr; one of the electrodes is rounded which act as anode and cathode is a hollow cylindrical cup. Arc discharge is initiated by touching the electrodes, once the discharge stabilizes; the electrodes are withdrawn to a distance of about 2mm. The inner diameter of cathode is 12 and length 30 mm. The diameter of anode is 15 and length 70 mm. High current flows through the electrodes and sublimation of graphite is ensued.

### 2.2 Cylindrical Hollow Cathode Direct Current Magnetron Sputtering

Fig 2 shows the schematic for CHC magnetron sputtering source enclosed in a glass tube evacuated to $10^{-6}$ Torr. Discharge is produced between a graphite hollow cathode and metal anode by applying 450 V and 40 mA with Neon as background gas at a pressure of 3-5 Torr. Cathode, which is also our sputtering target, is a 50 mm long hollow graphite cylinder with 17 mm inner diameter and wall thickness of 3 mm. Carbon is coated on Al and Cu substrates in the form of discs of 10 mm diameter, which act as anode and is placed inside the cathode at one third of the length. An axial magnetic field of 0.028 Tesla is applied which covers both cathode and anode but is perpendicular to anode and it helps to stabilize the discharge conditions to operate the source for longer time nearly one hour. During experiment, continuous flow of Neon gas is maintained to overcome the addition of impurities due to degassing.

### 2.3 Characterization Apparatus

The films produced are characterized using Raman and Fourier transform infrared (FTIR) spectroscopy, X-ray diffraction (XRD) and scanning electron microscopy (SEM). Raman spectroscopy is performed with R-3000 Raman system, USA using incident laser of wavelength 532 nm. Films are also characterized by FTIR spectroscopy (NICOLET 6700 by Thermoelectron Corporation, USA). XRD (D8 Discover made by Bruker axs, Germany) with Cu k-edge radiation was used to identify the crystallinity in the near surface region as a result of deposition. Scherrer's relation was used to estimate the grain sizes. The surface of coated substrates has been investigated by using SEM of model Leo-440i, UK, which characterizes the morphology of structures formed during the experiment.

## 2.4 Online Emission spectroscopy

State of sublimated and sputtered nanoparticles is analyzed using online emission spectroscopy. Level density $N_u$ of the upper excited levels can be obtained from emission spectrum by using the relation $I_{ul} = N_u A_{ul} h \upsilon_{ul}$ where $I_{ul}$ is the calibrated intensity ($\mu W/(cm^2$-nm)) obtained from the spectrum for relevant transition, $h\upsilon_{ul}$ is energy difference between upper ($u$) & lower ($l$) levels and $A_{ul}$ is Einstein transition probability of spontaneous emission for selected transition. Formula for evaluation of $N_u$ from the observed line intensity is valid for all spontaneous emissions [19]. Spectrum is background subtracted and calibrated against the experimental setup to get calibrated intensity $I_u$. Upper level density $N_u$ is used to calculate vibrational temperature ($T_{vib}$) by Boltzmann relation

$$N_u = \frac{g_u N}{U(T)} \exp\left(-\frac{E_u}{k_B T_{vib}}\right)$$

where $N$ and $N_u$ are total density of particles and density of particles in $u$ state respectively; $g_u$ is statistical weight of upper state; $U(T)$ is atomic or molecular internal partition function and $E_u$ is energy of upper level relative to ground state. From emission spectra if a range of upper level densities can be calculated, then by plotting $\ln\left(\dfrac{N_u}{g_u}\right)$ as a function of $E_u/k_B$ one can determine vibrational temperature ($T_{vib}$) from slope ($-1/T_{vib}$) [20]. Relative densities $N_u/N_l$ for two levels with respect to some reference level can be calculated corresponding to two calibrated intensities and from Boltzmann equation

$$\frac{N_u}{N_l} = \left(\frac{g_u}{g_l}\right) \exp\left[-\frac{(E_u - E_l)}{k T_{exc}}\right]$$

where $g_u$ and $g_l$ are the statistical weights, $E_u$ and $E_l$ are energies of the respective levels [19,21]. Excitation temperature $T_{exc}$ can be calculated from last relation for atomic or molecular species as the case may be. Emission spectroscopy is performed by Ocean optics spectrophotometers; one spectrometer is in visible (Vis) (400-600 nm) with resolution ~1 nm and other is ultraviolet-visible-near infrared (UV-Vis-NIR) (200-1100 nm) with higher resolution of 0.16 nm to especially identify $C_2$ Swan band $[d^3\prod_g -a^3\prod_u]$ [22].

## 3. Results and Discussion

Fig 3 shows Raman spectra of carbon deposited on Al and Cu samples for two sets of techniques. Fig 3a is for continuous arc discharge at current of 80 A whereas fig. 3b is for magnetron sputtering at discharge voltage of 450 V. A reference spectrum of polycrystalline graphite is displayed for comparison and two peaks at 1579 (G-band) and 1368 cm$^{-1}$ (D-band) are observed. Ratio of area under G and D-bands $I_D/I_G$ is 0.1 for pure graphite sample. From fig. 3a, intense G peak is observed at 1582 and relatively broader D peak at 1355 cm$^{-1}$. Area ratio $I_D/I_G$ for Al is 1.22 and for Cu is 1.56. From fig. 3b, a broader peak of G-band at 1580 and a small hump of D-band at 1365 cm$^{-1}$ is observed. $I_D/I_G$ is 0.42 for Al and 0.40 for Cu. In arc discharge, presence of both sp$^2$ and sp$^3$ bonding indicate more disorder as compare to magnetron sputtering. Film structures have more tendencies towards diamond like carbon. From $I_D/I_G$ for magnetron sputtering it is evident that films deposited are richer in sp$^2$ bonding and more inclined towards graphite like structures.

Fig. 4 gives FTIR spectra of carbon deposited on Al and Cu samples for two sets of techniques. Figures 4a and 4b are for arc discharge and magnetron sputtering respectively. In fig. 4a, the spectrum is dominated by two strong peaks at 2849 and 2920 cm$^{-1}$, which could be related to sp$^3$-CH$_3$ stretching modes and sp$^3$-CH$_2$ asymmetric modes, respectively. Sharp peak at 2360 cm$^{-1}$ is a signature of $CO_2$ present in atmosphere. Fig. 4b exhibits similar peaks for CO$^2$ and two minor peaks at 2860 and 2919 cm$^{-1}$ only for Cu substrate. Trace amount of hydrogen may be attributed to release of hydrogen from breaking of water molecules present in materials inside the vacuum chamber.

Fig. 5 depicts the X-ray diffraction patterns of carbon films from arc discharge. Figures 5a and 5b are for Cu and Al substrates respectively. From fig. 5a, substrate peaks are observed at 2θ, 43.3$^o$ (111), 50.4$^o$ (200) and 74.1$^o$ (220). The strongest peak of hexagonal graphite is observed at 26.5$^o$ (002). Another peak of hexagonal graphite at 43.4$^o$ (103) is shadowed under Cu broad peak. Grain size for hexagonal carbon is 55 nm. Two peaks of body centered cubic diamond are observed; one at 2θ, 48.5$^o$ (211) for the strongest intensity and second at 39.2$^o$ (200). Average precipitate grain size is calculated as 66 ± 17 nm. Peak at 51.6$^o$ (211) may correspond to orthorhombic diamond structure which is not the strongest intensity line. Grain size estimate comes out to be 88 nm. A peak of Rhombohedral diamond is observed at $75.3^{o}$ (110). The strongest peaks at $43.0^{o}$ (104) and $43.9^{o}$ (0115) are shadowed under Cu broad peak. Grain size for Rhombohedral diamond is 100 nm. It is obvious from X-ray diffraction pattern that both hexagonal carbon and diamond crystalline structures are formed as a result of high current arc discharge. From fig 5b, the strongest peak of rhombohedral $Al_4C_3$ is observed at $40.2^{o}$ (107). Grain size estimate comes out to be 169 nm. At $34.7^{o}$ (103), $37.8^{o}$ (104) and $65.0^{o}$ (116) peaks of hexagonal $Al_4C_3$ are observed. Average grain size is calculated as $82 \pm 30$ nm. Therefore, Al substrate has more tendency to form $Al_4C_3$. XRD graphs for magnetron sputtering showed only substrate peaks.

Fig. 6 display the scanning electron microscope (SEM) images of carbon films deposited on Cu and Al substrates with two techniques. Images in fig. 6 (a & b) have film thickness 3-4 µm by arc discharge whilst images in fig. 6 (c & d) have film thickness 6-7 µm by magnetron sputtering. From fig. 6 (a & b), films have rough and porous surfaces, which are not uniform in shape. From fig. 6 (c & d), films are relatively smooth and show micro-grains (carbon balls) which might be attributed to surface cracking. These grains are fairly round and seem to be agglomerates of carbon particles and similar in shapes. It is further noted that grains have uniform distribution throughout the surfaces. Cu surface has two to three different size of carbon balls as compared to Al surface, which has almost similar size of micro-grains. Fig. 7 represents the online emission spectra. Figures 7a and 7b are obtained from continuous arc discharge and magnetron sputtering respectively. Arc discharge shows clearly the molecular emission of C2 Swan band from which one can calculate $T_{vib}$ while CHC sputtering discharge has only CII atomic lines in addition to Neon. By applying Boltzmann graphical method $T_{vib}$ is calculated as $13600 \pm 480$ K for arc discharge and by applying ratio method excitation temperature $T_{exc}$ is calculated as $5600 \pm 380$ K for magnetron sputtering. For arc discharge from Raman spectra, film has more diamond like structures due to relatively higher temperature of subliming species, which enhances its $sp^3$ character. This fact is further supported by FTIR spectrum in which $sp^3$-$CH_3$ and $sp^3$-$CH_2$ modes are present. From XRD, Cu substrate shows some nanocrystalline structures whereas Al shows only carbides. From SEM images it is evident that surface seems to be a collection of loose agglomerates of carbon particles and are weakly bonded and has low adhesion with substrates. In contrast to arc discharge, the films by magnetron sputtering have different characteristics altogether. From Raman spectra, films have more $sp^2$ character may be due to relatively lower energies of dissociated carbon nanoparticles from cathode surface which become part of glow plasma as a result of sputtering. Deposited species may be neutral, metastable or negative carbon nanoparticles. These particles while settling down are bombarded by electrons and their motion is affected by axial magnetic field perpendicular to the substrate. Result of this bombardment of electrons is enhanced surface penetration and better mechanical adhesion. Absence of $sp^3$-$CH_3$ and $sp^3$-$CH_2$ modes from FTIR spectrum indicates unfavorable environment for hydrogen bonding. From XRD, Cu substrate shows no crystallinity verifying amorphous coatings.

## 4. Conclusions

This study is related to the technique of making amorphous carbon film in CHC magnetron sputtering from regenerative sooting discharge. Films are deposited on Cu and Al substrates and compared with carbon films deposited from conventional arc discharge between graphite electrodes.

In arc discharge Raman spectra reveals presence of both graphite and diamond like structures. This fact is further supported by FTIR spectrum in which $sp^3$-$CH_3$ and $CH_2$ modes are present whereas films deposited by magnetron sputtering have more tendencies towards $sp^2$ bonding. In XRD pattern some nanocrystalline structures are observed on Cu substrate in arc discharge and carbides are formed on Al substrate, whilst amorphous films are formed in magnetron sputtering. Emission spectrum of arc discharge confirms the presence of molecular species whereas atomic emission lines are observed in magnetron sputtering. SEM images depict loose agglomerates of carbon particles having rough and porous surfaces for arc discharge. In contrast the film for magnetron sputtering has compact, regular and round structures.

## Acknowledgements

PINSTECH provided financial support for this work from its R & D grant for scientific research. The authors gratefully acknowledge technical support provided by Mr. Rizwan, Mr. Faisal and GSD workshops for fabrication of different components.

**Figure Captions**

**Figure 1**: Experimental setup of continuous arc discharge technique with graphite cathode and anode to study the sublimation of graphite at discharge current of 80A.

**Figure 2**: Experimental setup of Magnetron Sputtering technique to coat carbon films on metallic substrates using Graphite Hollow Cathode.

**Figure 3**: Raman spectra of (a) Carbon soot collected at Cu and Al at current of 80 A in Arc discharge (b) Carbon coating on Cu and Al substrates in Hollow cathode magnetron sputtering.

**Figure 4**: FTIR spectrum of carbon films on Al and Cu substrates (a) in arc discharge at 80A (b) using hollow cathode magnetron sputtering.

**Figure 5**: XRD results samples coated with soot in arc discharge at 80A (a) Cu and (b) Al

**Figure6**: SEM micrographs of coated samples in arc discharge at 80A (a) Cu and (b) Al and with hollow cathode magnetron sputtering (c) Cu and (d) Al

**Figure 7**: Emission spectra (a) of $C_2$ Swan band system obtained from arc discharge at 80A

(b) obtained from hollow cathode Magnetron Sputtering

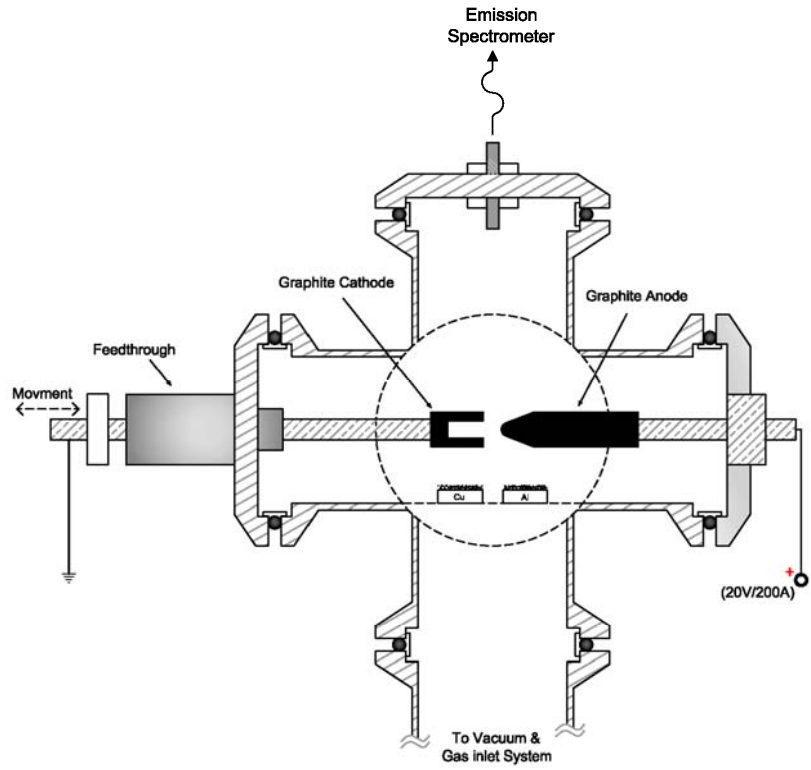

Figure 1

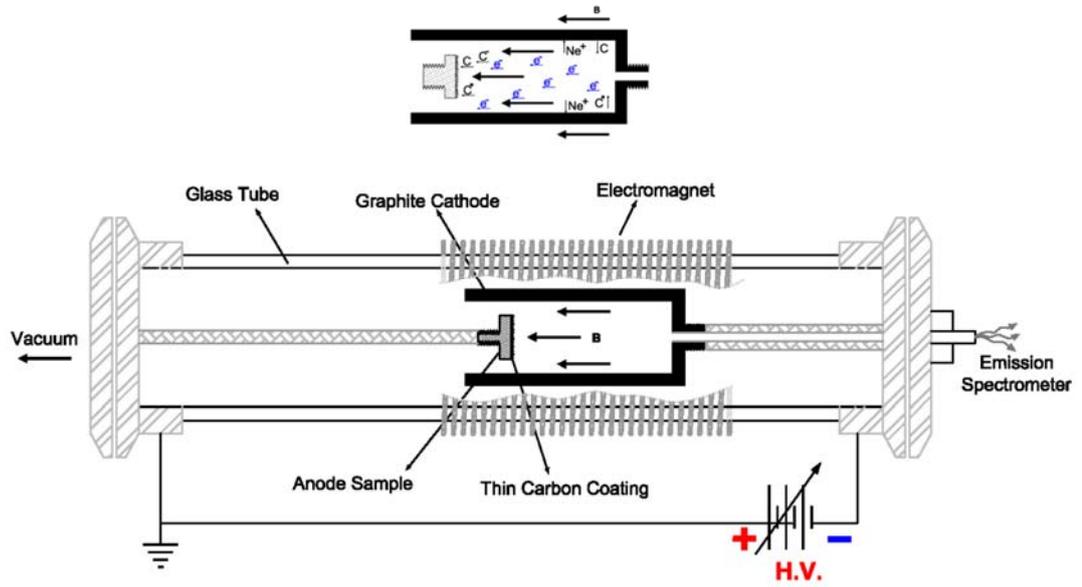

Figure 2

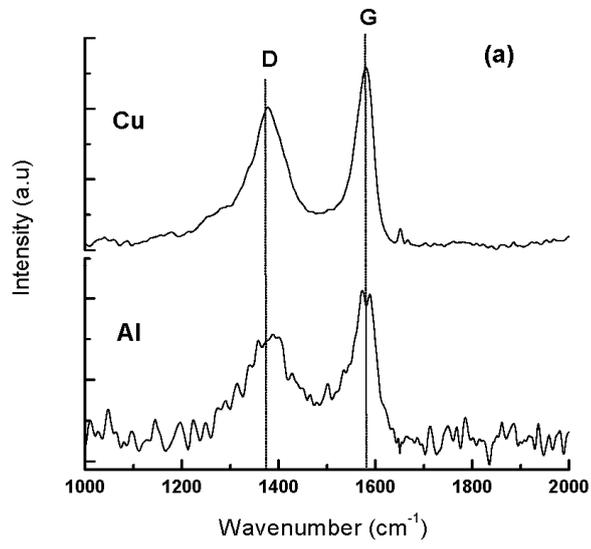

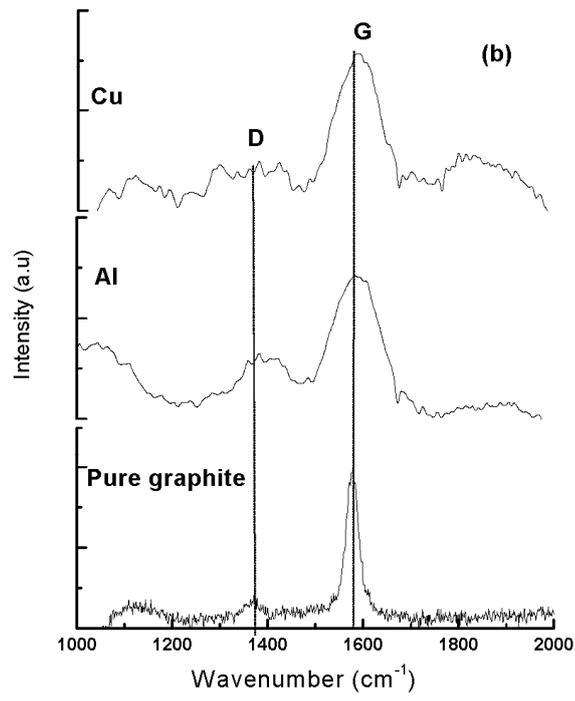

Figure 3

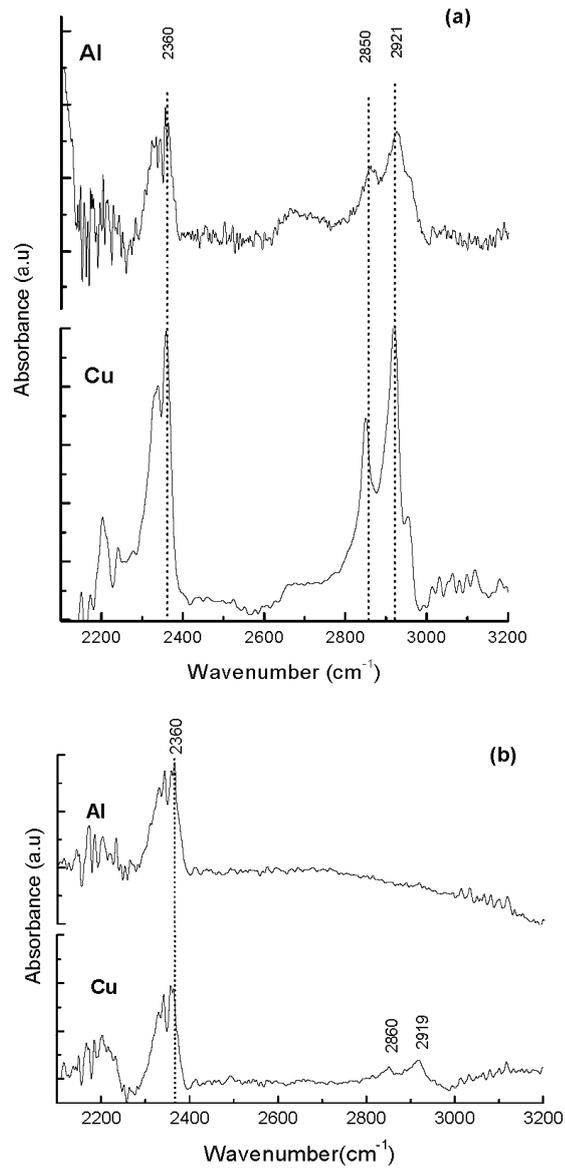

Figure 4

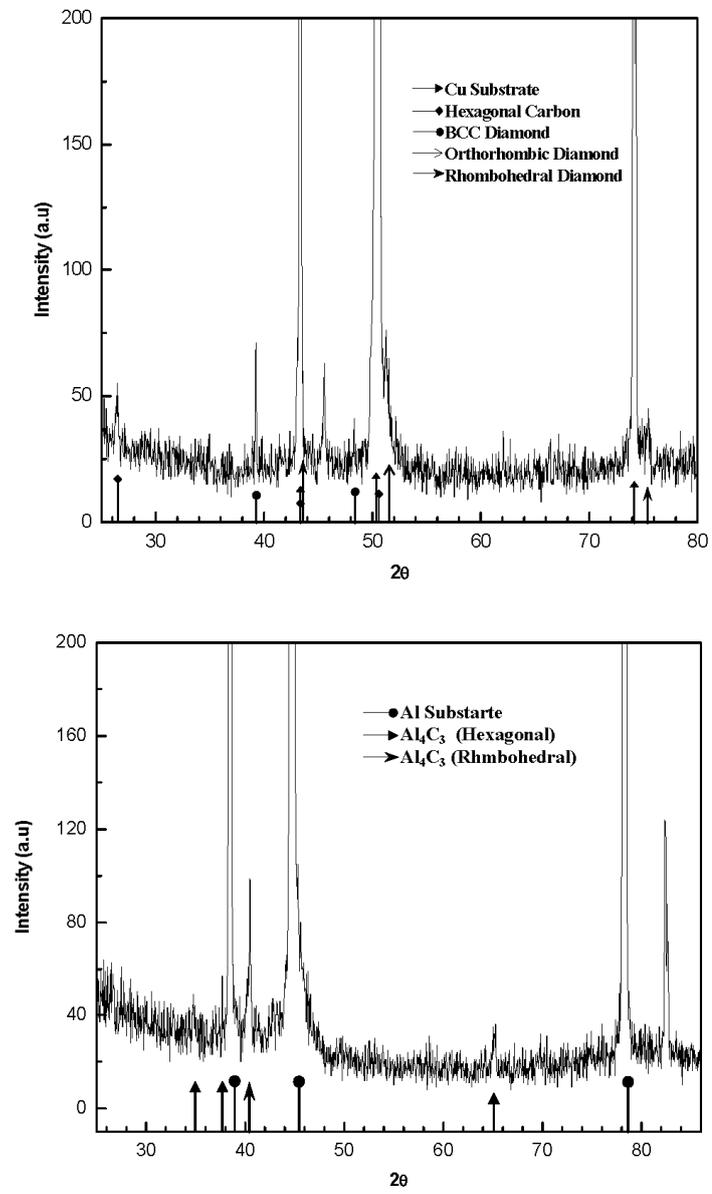

Figure 5

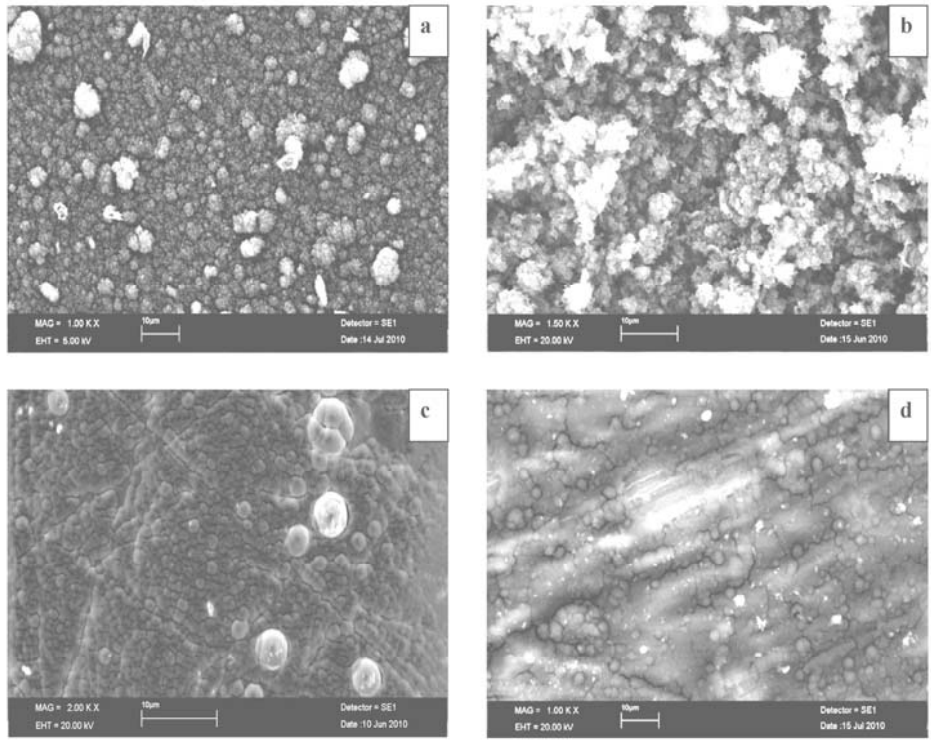

Figure 6

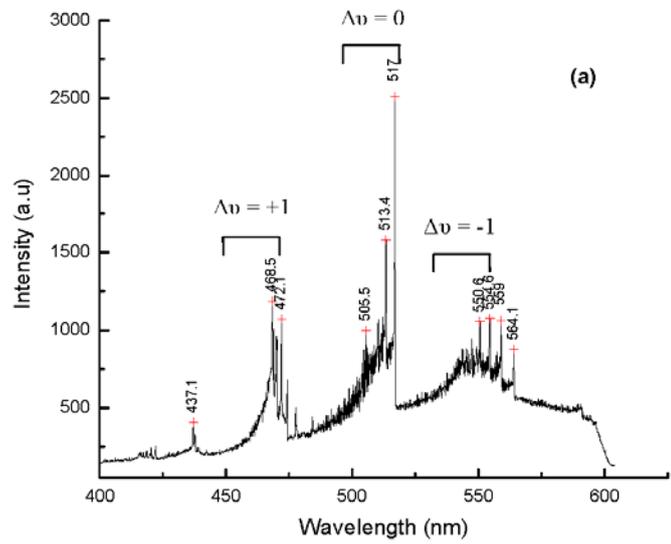

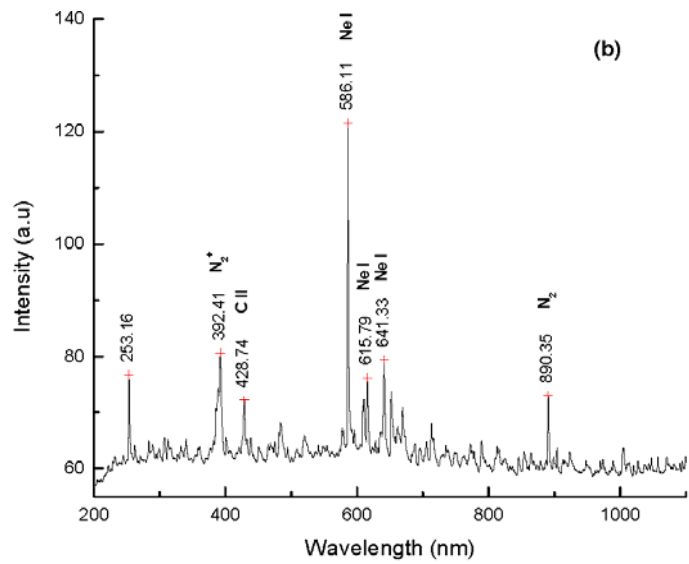

Figure 7